\documentclass[twocolumn,floatfix,aps,showpacs,showkeys]{revtex4}
\usepackage{epsfig,subfigure,latexsym}
\usepackage{color}
\usepackage{xcolor}
\usepackage[normalem]{ulem}
\usepackage{amssymb,amsmath}

\newcommand\redout{\bgroup\markoverwith
{\textcolor{red}{\rule[.5ex]{2pt}{0.4pt}}}\ULon}

\begin{document}
\def\Journal#1#2#3#4{{\it #1} {\bf #2}, #3 (#4) }
\def\RPP{{Rep. Prog. Phys}}
\def\PRC{{Phys. Rev. C}}
\def\PRD{{Phys. Rev. D}}
\def\ZPA{{Z. Phys. A}}
\def\NPA{{Nucl. Phys. A}} 
\def\JPG{{J. Phys. G }}
\def\PRL{{Phys. Rev. Lett}}
\def\PR{{Phys. Rep.}}
\def\PLB{{Phys. Lett. B}}
\def\AP{{Ann. Phys (N.Y.)}}
\def\EPJA{{Eur. Phys. J. A}}
\def\NP{{Nucl. Phys}}  
\def\RMP{{Rev. Mod. Phys}}
\def\IJMPE{{Int. J. Mod. Phys. E}}
\def\AJ{{Astrophys. J}}
\def\AJL{{Astrophys. J. Lett}}
\def\AA{{Astron. Astrophys}}
\def\ARAA{{Annu. Rev. Astron. Astrophys}}
\def\MPLA{{Mod. Phys. Lett. A}}
\def\ARNPS{{Annu. Rev. Nuc. Part. Sci}}
\def\LRR{{Living. Rev. Relativity}}

\title{Hyperons in neutron stars within Eddington-inspired Born-Infeld theory of gravity}

\author{A. I. Qauli, M. Iqbal, A. Sulaksono, and H. S. Ramadhan }

\affiliation{Departemen Fisika, FMIPA, Universitas Indonesia, Depok 16424, Indonesia }

\begin{abstract}
We investigate the mass-radius relation of neutron star (NS) with hyperons inside its core by using the Eddington-inspired Born-Infeld (EiBI) theory of gravity. The equation of state of the star is calculated by using the relativistic mean field model under which the standard SU(6) prescription and hyperons potential depths  are used to determine the hyperon coupling constants. We found that, for $4\times 10^{6}~\rm{m^2}~\lesssim~\kappa \lesssim~6\times 10^{6}~\rm{m^2}$, the corresponding NS mass and radius predicted by the EiBI theory of gravity is compatible with observational constraints of maximum NS mass and radius. The corresponding  $\kappa$ value is also compatible with the  $\kappa$ range  predicted by  the astrophysical-cosmological constraints. We also found that the parameter $\kappa$ could control the size and the compactness of a neutron star.
\end{abstract} 


\maketitle
 \section{INTRODUCTION}
\label{sec_intro}
Neutron stars with their extreme compactness and unknown composition make them a unique laboratory to investigate, not only strong gravitational field (see for examples Refs.~\cite{Will2009,Psaltis2008} for recent reviews), but also the equation of state (EOS) of extremely dense matter(see for examples Refs.~\cite{Lattimer2012,Chamel2013} for recent reviews). Based on a recent analysis on the mass distribution of a number of pulsars with secure mass measurement, it is found that $M$ $\sim$ 2.1 $M_\odot$ is an established lower bound value on maximum mass $(M_{max})$ for NS,  and the existence of more massive NSs with $M$ $\sim$ 2.5 $M_\odot$ is, in principle, possible ~\cite{KKYT2013}. The evidences of massive NS with accurate measurement, for example, are obtained from the recent observation of pulsar J1614-2230 from Shapiro delay \cite{Demorest10} with the mass $1.97~\pm~0.04~M_\odot$ and pulsar J0348+0432 from the gravitational redshift of its white dwarf companion \cite{Antoniadis13} with the mass $2.01~\pm~0.04~M_\odot$. In addition, there are also evidences that some black widow pulsars might have higher masses. For example, pulsar B1957+20 reportedly has a mass of $M$ = 2.4 $\pm$ 0.12 $M_\odot$~\cite{KBK2011}, and even gamma-ray black widow pulsar J1311-3430~\cite{RFSC2012} has higher mass but less accurate mass than  that of B1957+20. On the other hand, accurate measurements of the NS radii, if existing, would  also provide important information. Unfortunately, the analysis methods used to extract NS radii from observational data still have high uncertainty~\cite{MCM2013}. Furthermore, the limits of recent observational radii from different sources or even from the same source are often in contradictory to one another~\cite{Bog2013,Gui2013,LS2013,Leahy2011,Ozel2010,Steiner2010,Stein2013,Sule2011}. However, it is remarkable that a neutron star with radius $R_{1.4}$ = 10.7-13.1 km of canonical mass, is reported to be consistent with other observational analysis and the host of experimental data for finite nuclei~\cite{Steiner2010,LattimerLim}.

In many works, the mass and radius of neutron stars are usually used to constrain the equation of state of matter at high densities by assuming the general relativity (GR) theory as an ultimate theory of  gravitation. If accurate measurement of NS with the mass of greater than  2.4 $M_\odot$ is possible in the future, within GR this constraint means the EOS of the corresponding NS  should be very stiff. This fact is quite difficult to reconcile with possible existence of exotics such as hyperons in NS core and small measured radius that both favor soft EOS.  Whereas, all nuclear models that are compatible with the experimental data on hyper-nuclei predict the existence of hyperons in matter at the density of exceeding 2-3 times  nuclear saturation density ( $\rho_0= 0.16 \text{fm}^{-3}$) (see Refs.~\cite{Hyppuzz} and the references therein).  We need to note that up to now, there is no general agreement among the predicted results for the NS EOS by including hyperons and the maximum mass of the corresponding NS within GR framework. Even, in the last few years a lot of progress in this direction have been reported but many inconsistencies still remain. This problem is known as ``the hyperon puzzle'' (see Ref.~\cite{Lonardoni,Yamamoto,Artyom14} and the references therein).

However, the differences between GR and its alternatives or modifications become significant in the strong gravitational fields of neutron stars~\cite{DeDeo2003,Kazim2014}. Among the theories of gravity, a new kind of Eddington-inspired theory of gravity with Born-Infeld-like (EiBI) structure has been proposed by Banados and Ferreira~\cite{Banados10}. The EiBI theory of gravity shows distinctive features such as avoidance of singularities in the early cosmology and in the Newtonian collapse of presureless particles, the formation of stable pressureless stars, and the existence of pressureless cold dark matter with a non-zero Jeans length. However, the EiBI theory of gravity shows anomalies associated with the phase transition for negative $\kappa$ (see Ref.~\cite{Sham2013} and the references therein). It is also reported that the EiBI theory of gravity is safe from surface singularity pathology~\cite{Kim2014}. It is shown in Ref.~\cite{Sotani14} that the modified Tolman-Oppenheimer-Volkov (TOV) equation based on EiBI theory of gravity could adjust the maximum mass of NS by adjusting its $\kappa$ value, and the corresponding author has also found that through direct observations of the radii of low mass NS (around 0.5 $M_\odot$) and the measurements of neutron skin thickness of $^{208}$Pb, they could not only discriminate EiBI from GR but also estimate the  $\kappa$ value in EiBI. It is also reported that the range of  reasonable values of $\kappa$ parameter in EiBI model can be constrained by using some astrophysical and cosmological data~\cite{Avelino12}.

In this work, we demonstrate that ''hyperon puzzle''  problem that commonly appears if we use simple hyperon EOS within GR gravity is not present in EiBI theory of gravity. Note that to calculate the EOS of NS, we use the extended version of relativistic mean field (ERMF)model~\cite{SA2012,Furnstahl96,Furnstahl97}. We also found that it is possible to obtain NS with the mass of around  2.1 $M_\odot$ and the radius inside the range deduced by the authors of Ref.~\cite{Steiner2010} by using $\kappa$ value that is still compatible to the range obtained from astrophysical and cosmological constraints \cite{Avelino12}.  We need also to note that the models of NS for simple hyperon EOS with maximal mass  around  2.1 $M_\odot$ and within f(R) gravity has been studied in Ref.\cite{Artyom14} while within the anisotropic pressure assumption has been studied in Ref.~\cite{AS2014} . 

The paper is organized as follows. Sec.~\ref{sec_eos}, describes the brief outline of NS EOS. Sec.~\ref{sec_formalism} is devoted to discuss EiBI theory of gravity. Sec.~\ref{sec_ns} briefly describes the review of the derivation of TOV equation based on EiBI theory of gravity in NSs while Sec.~\ref{sec_ns} describes the numerical solutions and results. Finally, the conclusion is given in Sec.~\ref{sec_conclu}.

\begin{figure}
\epsfig{figure=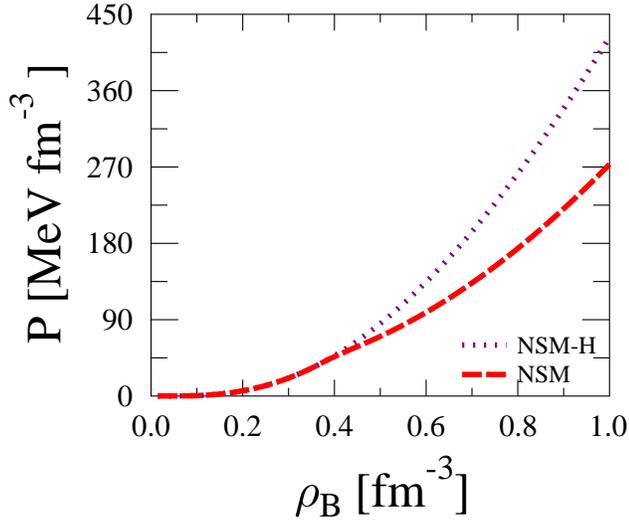, width=9.5cm}
\caption{EOS of neutron star matter based on the BSP parameter set of ERMF model with (NSM) and without hyperons (NSM-H).}
\label{fig:eos}
\end{figure}
\begin{figure}
\epsfig{figure=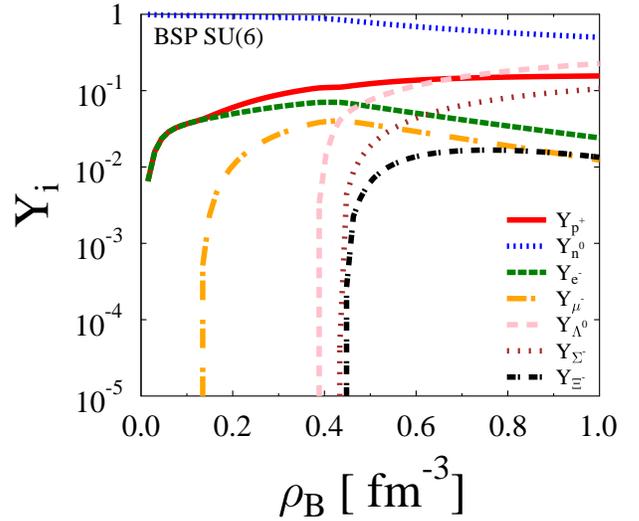, width=9.5cm}
\caption{Constituents fraction in neutron star core based on BSP parameter set of ERMF model.}
\label{fig:fraction}
\end{figure}

\section{EQUATION OF STATE}
\label{sec_eos}
NS can  be roughly divided into two regions with different compositions, particle distributions and density ranges namely the crust and core. In this work, we use the crust EOS calculated by Miyatsu $et ~al$.~\cite{MYN2013}, while the core is assumed to be composed of interacting baryons (nucleons and hyperons) and free leptons that are calculated using the ERMF model.  

The ERMF model is an extension of the standard RMF model by including additional cross-coupling terms for $\sigma$, $\omega$ and $\rho$ mesons~\cite{Furnstahl96,Furnstahl97}. In the RMF model, baryons interact each other by exchanging $\sigma$, $\omega$, $\rho$ and $\phi$ mesons. The total Lagrangian density can be written as~\cite{SA2012}
\begin{equation}
{\mathcal{L}} = {\mathcal{L}}^{\rm free}_{B} + {\mathcal{L}}^{\rm free}_{M} + 
  {\mathcal{L}}^{\rm lin}_{BM} + {\mathcal{L}}^{\rm nonlin}+{\mathcal{L}}^{\rm free}_{l},  
\label{Lag}
\end{equation}
where the free baryons Lagrangian density is,
\begin{equation}
{\mathcal{L}}^{\rm free}_{B}=\sum_{B=N,\Lambda,\Sigma,\Xi}\overline{\Psi}_B[i\gamma^{\mu}\partial_{\mu}-M_B]\Psi_B,
\end{equation}
Here,  $\Psi_B$ is baryons (nucleon, $\Lambda$, $\Sigma$ and $\Xi$) field. The Lagrangian density for the free mesons is, 
\begin{eqnarray}
{\mathcal{L}}^{\rm free}_{M}&=&\frac{1}{2}(\partial_{\mu}\sigma\partial^{\mu}\sigma-m_{\sigma}^2\sigma^2)+\frac{1}{2}(\partial_{\mu}\sigma^*\partial^{\mu}\sigma^*-m_{\sigma^*}^2\sigma^{*2})\nonumber\\ &-&\frac{1}{4}\omega_{\mu\nu}\omega^{\mu\nu}+\frac{1}{2}m_{\omega}^2\omega_{\mu}\omega^{\mu}-\frac{1}{4}\phi_{\mu\nu}\phi^{\mu\nu}+\frac{1}{2}m_{\phi}^2\phi_{\mu}\phi^{\mu}\nonumber\\ &-&\frac{1}{4}\mathbf{\rho}_{\mu\nu}\mathbf{\rho}^{\mu\nu}+\frac{1}{2}m_{\rho}^2\mathbf{\rho}_{\mu}\mathbf{\rho}^{\mu}.
\end{eqnarray}
The $\omega^{\mu\nu}$, $\phi^{\mu\nu}$ and $\mathbf{\rho}^{\mu\nu}$ are field tensors
corresponding to the $\omega$, $\phi$ and $\rho$ mesons field, and can be defined as
$\omega^{\mu\nu}=\partial^{\mu}\omega^{\nu}-\partial^{\nu}\omega^{\mu}$, $\phi^{\mu\nu}=\partial^{\mu}\phi^{\nu}-\partial^{\nu}\phi^{\mu}$
and $\mathbf{\rho}^{\mu\nu}=\partial^{\mu}\mathbf{\rho}^{\nu}-
\partial^{\nu}\mathbf{\rho}^{\mu}$, respectively. The Lagrangian ${\mathcal{L}}^{\rm lin}_{BM}$ describing interactions among baryons through mesons exchange can be written as 
\begin{eqnarray}                   
{\mathcal{L}}^{\rm lin}_{BM}&=& \sum_{B=N,\Lambda,\Sigma,\Xi}\overline{\Psi}_B[g_{\sigma B} \sigma + g_{\sigma^* B} \sigma^* -\gamma_\mu g_{\omega B} \omega^\mu\nonumber\\&-&\frac{1}{2}\gamma_\mu g_{\rho B}\mathbf{\tau_B}\cdot \mathbf{\rho} ^\mu -\gamma_\mu g_{\phi B}\phi ^\mu ]\Psi_B,
\label{eq:Llin}
\end{eqnarray}
where $\tau_B$ is the baryons isospin matrices. The Lagrangian describing
mesons self interactions for $\sigma$, $\omega$, and $\rho$ is defined as,
\begin{eqnarray}
 {\mathcal{L}}^{\rm nonlin} &=& - \frac{\kappa_3 g_{\sigma N} m_{\sigma}^2}{6 m_{ N} } \sigma^{3}
                   - \frac{\kappa_4 g_{\sigma N}^2 m_{\sigma}^2}{24 m_{ N}^2 } \sigma^{4}+\frac{\zeta_0 g_{\omega N}^2}{24} 
                   {(\omega_{\mu}  \omega^{\mu})}^2\nonumber\\
&+& \frac{\eta_1 g_{\sigma N} m_{\omega}^2}{2 m_{ N} } \sigma \omega_{\mu}  \omega^{\mu}+\frac{\eta_2 g_{\sigma N}^2 m_{\omega}^2}{4 m_{ N}^2 }\sigma^{2} \omega_{\mu}  \omega^{\mu} 
\nonumber\\&+&\frac{\eta_{\rho} g_{\sigma N} m_{\rho}^2}{2 m_{ B} } \sigma
\mathbf{\rho}_{\mu} \cdot \mathbf{\rho}^{\mu} +\frac{\eta_{1\rho}g_{\sigma
N}^2m_{\rho }^{2}}{4m_N^2} \sigma^2\mathbf{\rho}_{\mu} \cdot \mathbf{\rho}^{\mu} \nonumber\\ &+&\frac{\eta_{2 \rho} g_{\omega N}^2 m_{\rho}^2}{4 m_{ N}^2 } \omega_{\mu}  \omega^{\mu} \mathbf{\rho}_{\mu} \cdot \mathbf{\rho}^{\mu}.
\label{eq:Lnlin}
\end{eqnarray}
 While the free leptons Lagrangian density is, 
\begin{equation}
{\mathcal{L}}^{\rm free}_{l}=\sum_{l=e^-, \mu^-}\overline{\Psi}_l[i\gamma^{\mu}\partial_{\mu}-M_l]\Psi_l.
\end{equation}

here $\Psi_l$ is the leptons (electron and muon) field. The nucleons coupling constant and nonlinear parameters (BSP parameter set) can be found in Ref.~\cite{SA2012}. The vector part of hyperons coupling constant $g_{\omega H}$ and $g_{\phi H}$ can be obtained from standard prescription based on SU(6) symmetry~\cite{JSB_AG} namely

 \begin{eqnarray}
\frac{1}{3}g_{\omega N}&=&\frac{1}{2}g_{\omega
\Lambda}=\frac{1}{2}g_{\omega \Sigma}=g_{\omega \Xi},\nonumber\\
g_{\rho N}&=&\frac{1}{2}g_{\rho \Sigma}=g_{\rho \Xi},~ ~ ~ ~ ~
~ g_{\rho\Lambda}=0, \nonumber\\ 2 g_{\phi \Lambda}&=&2 g_{\phi
\Sigma}=g_{\phi \Xi}=\frac{2 \sqrt{2}}{3}g_{\omega N},~ ~ ~ ~ ~ ~
g_{\phi  N}=0.  
\label{eq:su6}
\end{eqnarray} 

For the given values of $g_{\omega H}$, the scalar hyperons coupling strengths  $g_{\sigma H}$ are obtained from the hyperons potential depth in the symmetric nuclear matter that is evaluated at the saturation density $\rho_0$ as,
\begin{equation} 
U_{H}^{(N)}(\rho_0)
= -g_{\sigma H}\sigma(\rho_0)+g_{\omega H}\omega(\rho_0).
  \label{eq:uyn} 
\end{equation}
The experimental values of potential depth $U_H^{(N)}$ at  $\rho_0$ are taken from Ref.~\cite{JSB_AG} namely
\begin{eqnarray}
 U_{\Lambda}^{(N)} &=& -28 {\rm ~MeV}, \quad U_{\Sigma}^{(N)} = +30 {\rm ~MeV}  \nonumber\\
&{\rm and}&  \quad U_{\Xi}^{(N)} = -18 { \rm ~MeV}.  
\label{eq:depth}
 \end{eqnarray}

The constituents composition in NS core should obey the chemical potential balance, charge neutrality and baryon density conservation ($\beta$ stability) conditions. Once the  momentum Fermi of every constituent involved is known from $\beta$ stability conditions, the total energy density ($\epsilon$) of NS core matter which is equivalent to the zero component of energy-momentum tensor ($T^{00}$), can be calculated numerically from Eq.~(\ref{Lag}) by using the standard procedure of mean field approximation. Detailed procedure  of $\epsilon$ derivation in mean field approximation for examples can be found in standard text books such as Refs.~\cite{Glendenning,Walecka}. While the radial pressure $P$ can be obtained in general from the vector component of energy-momentum tensor ($T^{ii}$) or can be calculated numerically from  thermodynamic relation i.e.,
\begin{equation}
P = \rho_B^2 \frac {d(\epsilon/\rho_B)}{d\rho_B}, 
\end{equation}
where $\rho_B$ is baryon density. The effect of including hyperons in neutron star matter can be seen in Fig.~\ref{fig:eos} while the core constituents fraction for neutron star with hyperons in it, is shown in Fig.~\ref{fig:fraction}. It is obvious from Fig. ~\ref{fig:eos} that EOS of NS matter with hyperons becomes softer starting from $\rho_B$ $\approx$ 0.4 compared to the one without hyperons because it can be seen in Fig.~\ref{fig:fraction}, for  $\rho_B$  $\ge$ 0.4 that corresponding to $P$ $\ge$ 50 MeV fm$^{-3}$ of NS matter, slow moving $\Lambda, \Sigma^-, \Xi^-$ hyperons start to appear and the number of energetic nucleons and leptons decreases. The EOS, or explicit  $\epsilon$($P$) relation will be used as input to solve the Tolman-Oppenheimer-Volkoff (TOV) equations for the EiBI theory of gravity in Sec.~\ref{sec_num}.

\section{EDDINGTON-INSPIRED BORN-INFELD THEORY OF GRAVITY}
\label{sec_formalism}

In this section, we will briefly discuss the EiBI formalism proposed by Banados and Ferreira\cite{Banados10} to describe compact stars. The Eddington-inspired Born-Infeld gravity theory is a subclass within the nonlinear theory of gravity. It is based on the nonlinear theory of electrodynamics, known as the Born-Infeld theory~\cite{Born:1934gh}. Banados and Ferreira~\cite{Banados10} later proposed a nonlinear theory of gravity having a Born-Infeld structure. The action of  EiBI theory of gravity is given by
\begin{eqnarray}
\nonumber  S &=&\frac{1}{8\pi\kappa} \int d^4x \left({\sqrt{-|g_{\mu\nu}+\kappa R_{\mu\nu}|} - \lambda \sqrt{-g}}\right)\\
&& + S_M[g,\Psi_M], \label{a}
\end{eqnarray}
where  $R_{\mu\nu}$ is symmetric Ricci tensor. In the EiBI theory of gravity, this tensor is constructed in the Palatini formulation. Therefore, $R_{\mu\nu}$ is a functional of  connection $\Gamma^\alpha_{\mu\nu}$, $R[\Gamma]$. Meanwhile, $\kappa$ and  $\lambda$ are parameters that are related to the Born-Infeld non-linearity and the cosmological constant, respectively. If $\kappa$ is going to zero, Eq.(\ref{a}) will reduce to the action of standard GR gravity. Here $|g_{\mu\nu} + \kappa R_{\mu\nu}|$ denotes the absolute value of the determinant of the tensor $(g_{\mu\nu} + \kappa R_{\mu\nu})$.\\
By varying the action in Eq. (\ref{a}) \cite{Harko13,Banados10,Delsate12,Vollick04},we can obtain the following equations: 
\begin{eqnarray}
q_{\mu\nu} &=& g_{\mu\nu} + \kappa R_{\mu\nu} \label{b}\\
q^{\mu\nu} &=& \tau \left(g^{\mu\nu} - 8 \pi \kappa  T^{\mu\nu} \right) \label{c}\\
\Gamma^\alpha_{\beta\gamma} &=& \frac{1}{2} q^{\alpha\rho}(q_{\rho \beta,\gamma} + q_{\rho \gamma,\beta} - q_{ \beta \gamma,\rho}), \label{d}
\end{eqnarray}
where $q_{\mu\nu}$ is an auxiliary metric, $\tau \equiv \sqrt{g/q}$, and $q$ is the determinant of metric $q_{\mu\nu}$.

From Eqs. (\ref{b}) and (\ref{c}), one can find mixed Einstein tensor $G^\mu_\nu$ i.e. \cite{Delsate12},
\begin{eqnarray}
\nonumber G^\mu_\nu &\equiv& R^\mu_\nu - \frac{1}{2} R \delta^\mu_\nu\\
&=& 8\pi\tau T^\mu_\nu - \left({\frac{1-\tau}{\kappa} + 4\pi\tau T}\right)\delta^\mu_\nu. \label{e}
\end{eqnarray}
We note that $G^\mu_\nu$ and $R^\mu_\nu$, are defined in terms of auxiliary metric. The factor $\tau$ can be obtained by multiplying the Eq. (\ref{c}) by metric $g_{\nu\alpha}$ and then taking its determinant. The explicit form of $\tau$ is \cite{Delsate12}
\begin{eqnarray}
\tau = {|\left({\delta^\mu_\nu - 8 \pi \kappa T^\mu_\nu}\right)|}^{-\frac{1}{2}}.
\end{eqnarray}

It is worthy to note from the coupling-to-the-matter perspective, EiBI can be considered as GR with additional isotropic pressure in apparent stress tensor. This pressure depends on $\tau$. In this view, $\tau$  plays crucial role in determining the corresponding apparent EOS. On the other hand, from observations of mass-radius relations or quasi-normal mode frequencies of neutron stars, gravitational waves in general etc, one may only obtain information about the apparent EOS. Furthermore, the authors of Ref.\cite{Delsate12} also argued that infinite  $\tau$ for finite values of $\epsilon$ and $P$ triggers singularity avoidance (see Ref.\cite{Delsate12} for the details of the $\tau$  significance ).

\section{NEUTRON STARS IN EiBI THEORY OF GRAVITY}
\label{sec_ns}

\begin{figure*}
\epsfig{figure=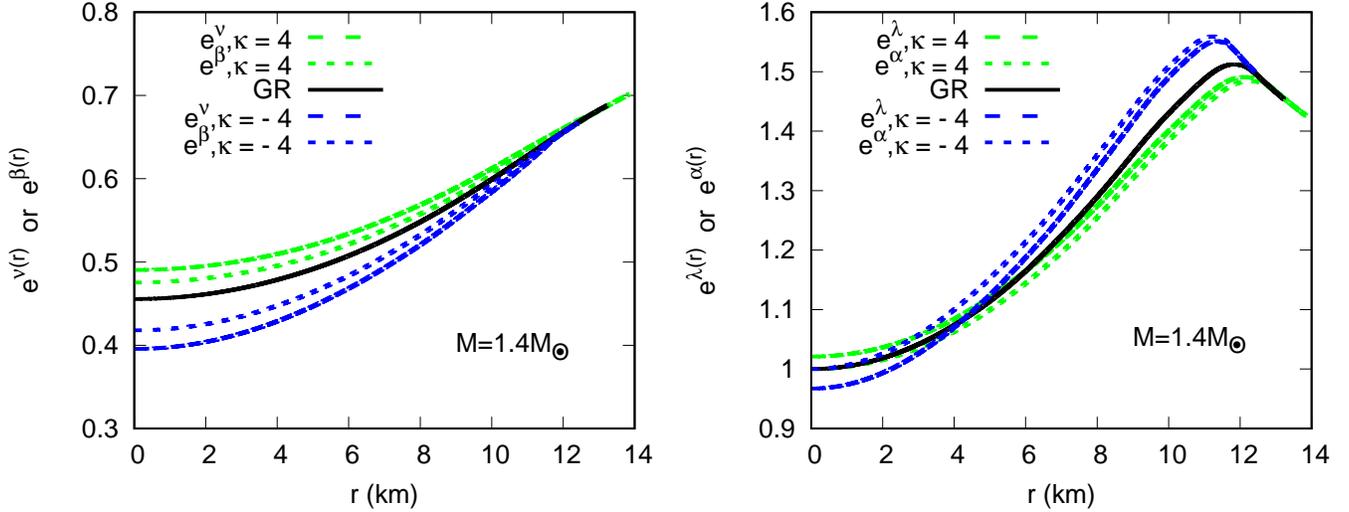, width=18cm}
\caption{ Profiles of exponential of metrics $e^\nu$, $e^\beta$ (left panel) and  $e^\lambda$, $e^\alpha$ (right panel) as a function radius coordinate r  for $\kappa$=4,-4 and 0 (GR). The black solid line represents the result of GR where in this case $e^\nu$= $e^\beta$ and $e^\lambda$= $e^\alpha$. These plots are taken for NS with mass $M=1.4~M_\odot$ case. Note that $\kappa$ is in unit 10$^6$ m$^2$.}
\label{fig:metric}
\end{figure*}

As it is mentioned in the introduction that many studies of NS properties used GR as the ultimate theory of gravity. Even though a lot of progress are reported in this direction, but until now the core EOS is still uncertain and the problem such as hyperon puzzle still remains.  On the other hand, NS is a strong gravitational object where the differences between GR and alternative or modified gravity theories such as EiBI can be significant\cite{DeDeo2003,Kazim2014}. However, only recently several authors applied the EiBI theory of gravity to study compact objects~\cite{Harko13,Banados10,Delsate12,Sotani14}. Therefore, it is still useful for the readers if we briefly review the derivation of TOV based on EiBI theory of gravity.

We start from standard assumption that the EOS of NSs satisfies the energy-momentum tensor of perfect fluid, i.e.,
\begin{eqnarray}
T_{\mu\nu} = (\epsilon + p)u_\mu u_\nu + p g_{\mu\nu},
\label{tmn}
\end{eqnarray}
which satisfies the conservation equation, $\nabla_\mu T^{\mu\nu}=0$. In Eq.~(\ref{tmn}), $\epsilon$, $p$, and $u_\mu$ denote the energy density, the isotropic pressure, and the four velocity of the NS matter, respectively. Now we introduce the line element of the metric $g_{\mu\nu}$ and the auxiliary metric $q_{\mu\nu}$ that describe the structure of compact static and spherically symmetric objects \cite{Sham2013,Harko13} 
\begin{eqnarray}
g_{\mu\nu}dx^\mu dx^\nu = -e^{\nu(r)} c^2dt^2 + e^{\lambda(r)}dr^2 + f(r) d\Omega^2\nonumber\\
q_{\mu\nu}dx^\mu dx^\nu = -e^{\beta(r)} c^2dt^2 + e^{\alpha(r)}dr^2 + r^2 d\Omega^2.
\end{eqnarray}
By using these definition for functions $a$ and $b$ as
\begin{eqnarray}
a \equiv \sqrt{1+\frac{8\pi G \kappa\epsilon }{c^2}}\label{eq:ab1}\\
b \equiv \sqrt{1-\frac{8\pi G \kappa p }{c^4}},
\label{eq:ab2}
\end{eqnarray}
and finding the $tt$ and $rr$-components of Eq. (\ref{e}) we will obtain these following equations
\begin{eqnarray}
\frac{d}{dr} \left({r e^{-\alpha}}\right) = 1 - \frac{1}{2\kappa} \left({2 + \frac{a}{b^3} - \frac{3}{ab}}\right)r^2, \label{g}\\
e^{-\alpha} \left({1 + r \beta'}\right) =  1 + \frac{1}{2\kappa} \left({\frac{1}{ab} + \frac{a}{b^3} - 2}\right)r^2,\label{j}
\end{eqnarray}
where the prime sign in $\beta$ variable in Eq.~(\ref{j}) and the other variables in the remaining equations in this work means the first derivative of the corresponding variable in respect of $r$. From Eq.(\ref{c}) we can obtain the following relations
\begin{eqnarray}
e^{\beta} = \frac{e^\nu b^3}{a},~~~ e^\alpha = e^\lambda ab,~~~ f = \frac{r^2}{ab}.\label{f}
\end{eqnarray}
\begin{figure*}
\epsfig{figure=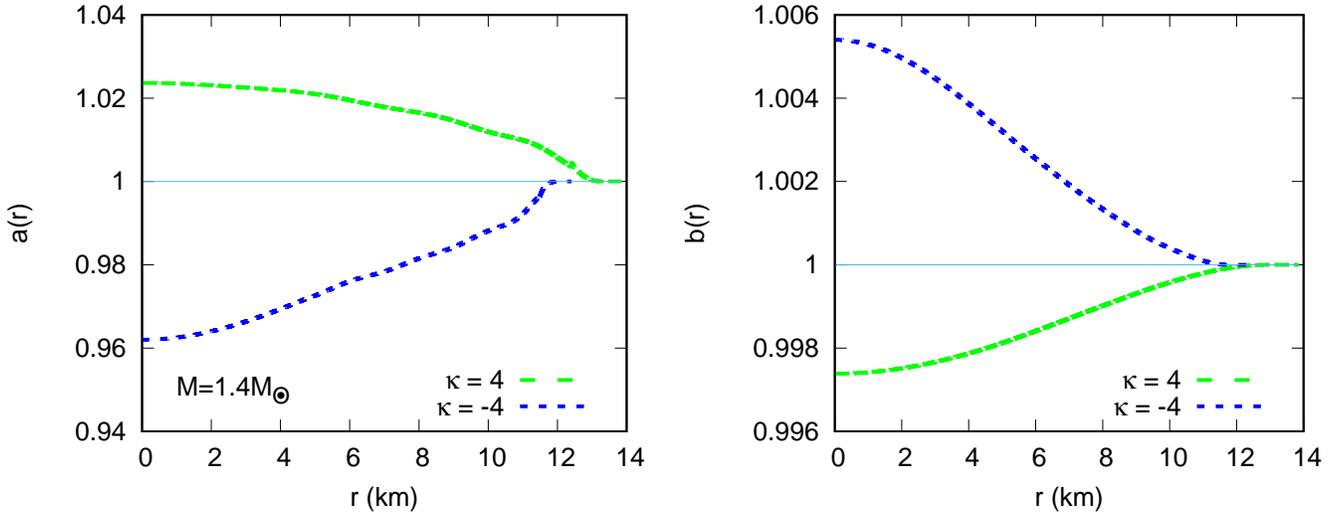, width=18cm}
\caption{Profiles of parameter $a$ and $b$  in Eq.~(\ref{eq:ab1}) and (\ref{eq:ab2}) as a function of radius coordinate r. $\kappa$ is in unit 10$^6$ m$^2$.}
\label{fig:metric_f}
\end{figure*}

On the other hand, from the conservation of energy-momentum in the $g$-metric we can obtain
\begin{eqnarray}
\nu' = \frac{4b}{a^2-b^2} b'.\label{nu}
\end{eqnarray}
So that  $\beta'$ in the first equality of Eq. (\ref{f}) can be written as
\begin{eqnarray}
\beta' = \frac{4b}{a^2-b^2}b' + \frac{3}{b}b' - \frac{1}{a}b'.
\end{eqnarray}
By defining the speed of sound $c_q^2 = (\frac{da(b)}{db})$ =$(\frac{dp}{d\epsilon})$,  $\beta'$ becomes
\begin{eqnarray}
\beta' = \left({\frac{4b}{a^2-b^2} + \frac{3}{b} - \frac{1}{a}c^2_q}\right)b'.\label{h}
\end{eqnarray}
One can easily integrate Eq. (\ref{g}) and the result is
\begin{eqnarray}
e^{-\alpha} = 1 - \frac{2Gm(r)}{c^2r},\label{i}
\end{eqnarray}
where 
\begin{eqnarray}
m' = \frac{c^2}{4G\kappa} \left({2 + \frac{a}{b^3} - \frac{3}{ab}}\right)r^2.
\label{TOVa}
\end{eqnarray}

The similar form as the standard TOV equation pressure derivative can be obtained by substituting Eqs. (\ref{h}) and (\ref{i})  into Eq. (\ref{j}).  The result can be written as 
\begin{eqnarray}
p' = -\frac{bc^4}{4 \pi G \kappa}\frac{ab(a^2 - b^2)[\frac{1}{2\kappa}(\frac{1}{ab}+\frac{a}{b^3}-2)r^3 + \frac{2Gm}{c^2}]}{r^2\left({1-\frac{2Gm}{c^2r}}\right)[4ab^2 + (3a - bc_q^2)(a^2 - b^2)]}.\nonumber\\
\label{TOV}
\end{eqnarray}
For more details regarding the derivation of Eq. (\ref{TOV}), one can consult Ref. \cite{Harko13}.  We also need to note that  EiBI  and GR theories are identical on the region outside the star (r $\ge$ $R$). Therefore, we can use the same boundary conditions  at $r=R$ as those of GR not only for solving TOV but also the metric  $e^{\nu}$ equations. 

In conclusion, we can obtain static NS properties based on the EiBI theory of gravity by explicitly solving Eqs.~(\ref{TOVa}) and (\ref{TOV}).

\begin{figure}
\epsfig{figure=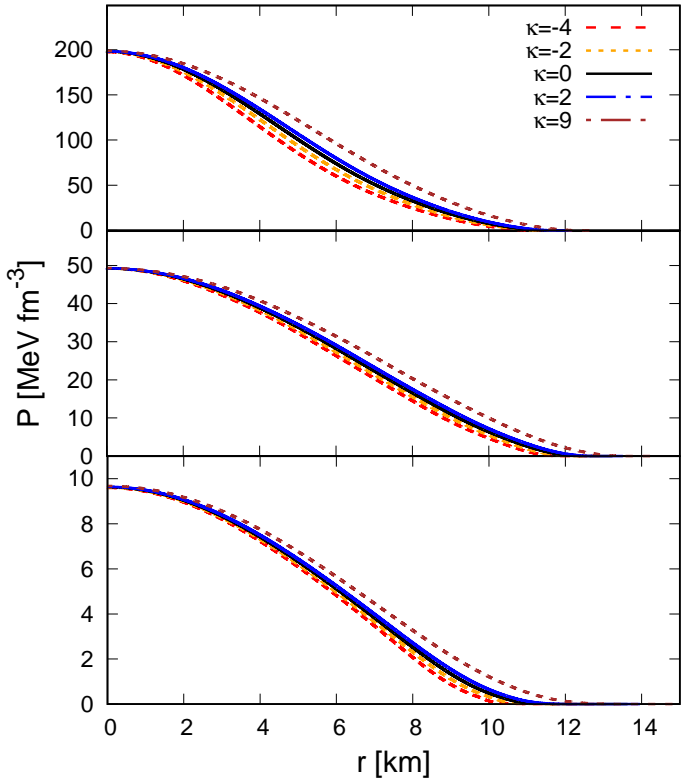, width=9.5cm}
\caption{Pressure profile as a function radius coordinate r for a NS with several $\kappa$ values. The upper panel is for a NS with $P_c$ = 200 $\rm MeV~fm^{-3}$, middle panel for NS with $P_c$ = 50 $\rm MeV ~fm^{-3}$ and lower panel for NS with  $P_c$ = 10 $\rm MeV fm^{-3}$. $\kappa$ is in unit 10$^6$ m$^2$.}
\label{fig:pressure}
\end{figure}

\begin{figure}
\epsfig{figure=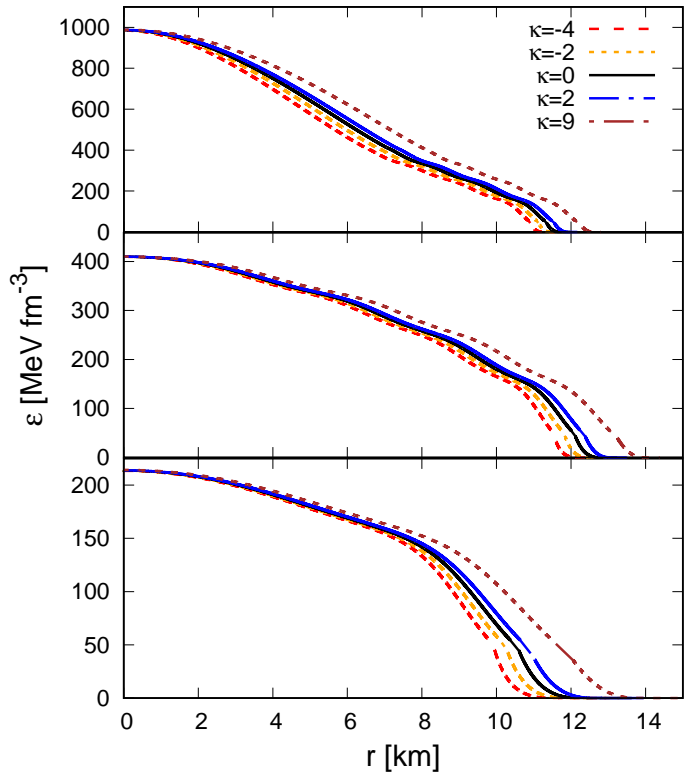, width=9.5cm}
\caption{Energy profile as  a function radius coordinate r for a NS with several $\kappa$ values. Each panel uses exactly the same central pressure  as those used in Fig~\ref{fig:pressure}. $\kappa$ is in unit 10$^6$ m$^2$.}
\label{fig:eden}
\end{figure}

\section{Numerical solutions and results}
\label{sec_num}
\subsection{Numerical methods and goals}
\begin{figure}
\epsfig{figure=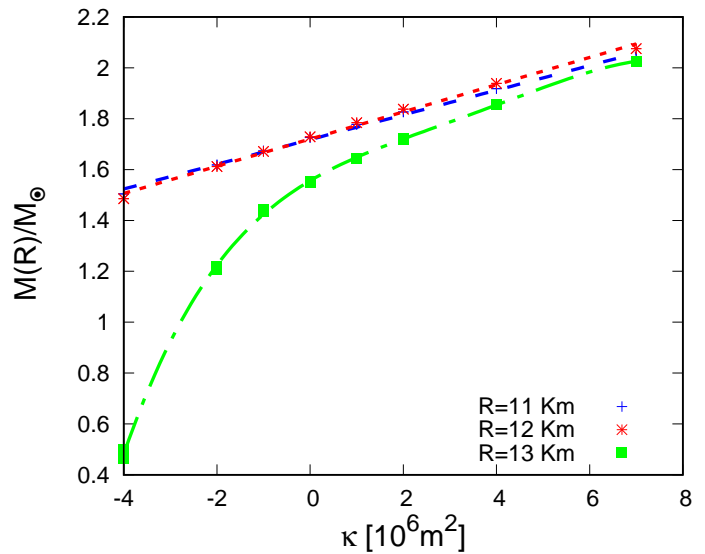, width=9.5cm}
\caption{The relationship between NS $M/M_{\odot}$ and $\kappa$ for some particular fixed $R$.}
\label{fig:M-K}
\end{figure}

\begin{figure}
\epsfig{figure=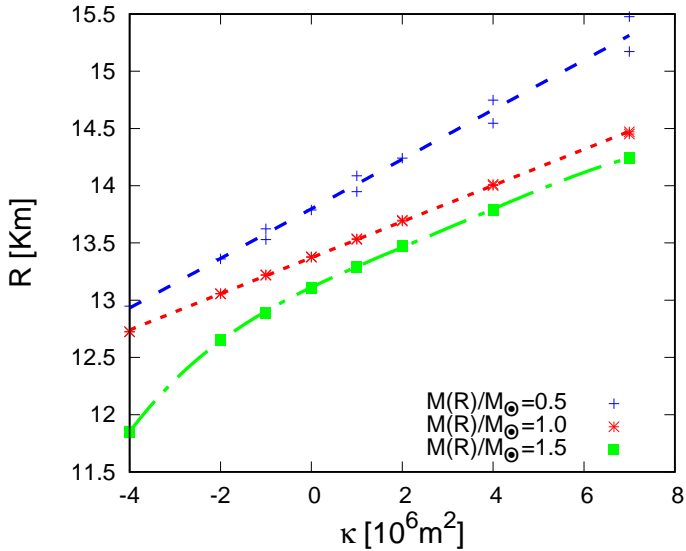, width=9.5cm}
\caption{The relationship between NS radius and $\kappa$ for particular fixed NS $M/M_{\odot}$.}
\label{fig:R-K}
\end{figure}

Principally, once the EOS of the corresponding star and $\kappa$ value of EiBi theory of gravity model are given, TOV equations (Eqs. (\ref{TOVa}) and (\ref{TOV})) can be integrated numerically by using fourth order Runge-Kutta algorithm starting from central  $r_{\epsilon}$ $\sim$ 0 until the edge of the star ($r=R$). The initial conditions  in the center of the star to solve these equations are given i.e., $P(r_{\epsilon})=P_c$, $m(r_{\epsilon})\sim 0$ and  the radius of the star $R$ is determined from the condition $P($R$)\sim 0$. If the latter condition is fulfilled then $m(R)= M$. In this way, for every given central pressure $P_c$ value, we can calculate its corresponding mass $M$, radius $R$, compactness $\xi$, and redshift $z$ of the star as well as their corresponding pressure, energy density and mass profiles. 

Similarly, the solution for metric $\nu$ profile can also be obtained by solving simultaneously  Eqs.~(\ref{nu}), ~(\ref{TOVa}), and (\ref{TOV}). However, different from those of Eqs.~(\ref{TOVa}) and ~(\ref{TOV}) where the  mass and pressure values at $r=r_\epsilon$ are known, for the metric $\nu$, the only information is $\nu$ value at $r=R$ ($e^{\nu(R)}=1-\frac{2GM}{c^2R}$). Therefore, in this case, we simply guess some particular value of  $\nu$ at $r=r_\epsilon$ and repeat the Runge-Kutta calculation several times. If the  $\nu$ value at $r=R$ fulfills the required  value by the corresponding boundary condition, hence it means our used $\nu$ value at $r=r_\epsilon$ is the correct one. In this way, we can calculate the $e^\nu$ profile. Other metrics profiles such as  $e^\alpha$, $e^\beta$ and $e^\lambda$ can be calculated by using $e^\nu$, pressure, energy density and mass profiles. 

\begin{figure}
\epsfig{figure=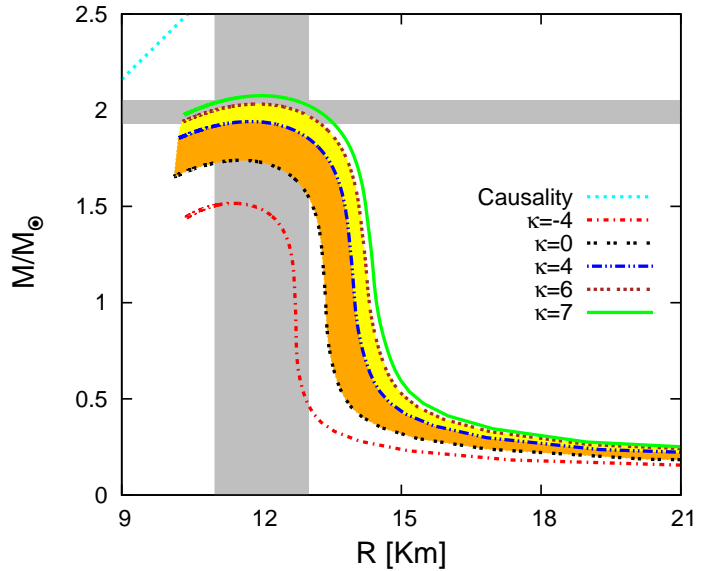, width=9.5cm}
\caption{NS mass-radius relation for some values of $\kappa$ where  $\kappa$=0 corresponds to GR result. The horizontal shaded band is the pulsar mass constraint from Ref.\cite{Antoniadis13} and the vertical shaded band is the radius constraint from  Ref.~\cite{Steiner2010}. The combination of yellow and orange shaded bands is from astrophysical-cosmological constraints \cite{Avelino12}. The constraint for $\kappa$ we get in this work, that is yellow shaded band only, is deducted both from astrophysical-cosmological constraint and maximum mass constraint in horizontal shaded band where $4~\lesssim~\kappa \lesssim~6$. The dot line represents the limit of causality of GR from \cite{LP2007}. Note that $\kappa$ is in unit 10$^6$ m$^2$. }
\label{fig:radmass}
\end{figure}

\begin{figure}
\epsfig{figure=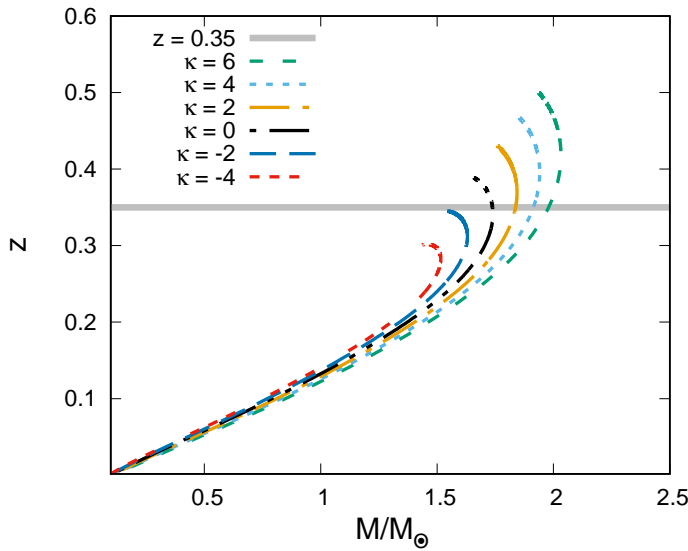, width=9.5cm}
\caption{Redshift $z$ as a function of $M/M_{\odot}$. The horizontal line is $z$ for low mass X-ray binary EXO 0748-676 NS~\cite{Cottam02}. $\kappa$ is in unit 10$^6$ m$^2$.}
\label{fig:redshift1}
\end{figure}

The main numerical results will be shown in Figs.~\ref{fig:metric}-\ref{fig:compactness} and they will be discussed in next subsections.

\subsection{Numerical solutions for various quantities and for different $\kappa$ values}
In this part, we discuss the consequences of applying the EiBI theory of gravity to describe NSs properties. 

In Fig.~\ref{fig:metric}, the  profiles of exponential of metrics $e^\nu$ and its auxiliary counter part $e^\beta$ (left panel) as well as  $e^\lambda$ and  its auxiliary counter part $e^\alpha$ (right panel) for NS with $M=1.4~M_\odot$  in the cases $\kappa$=-4~10$^6$ m$^2$ and 4~10$^6$ m$^2$ are given. The black solid line represents the result of GR ($\kappa$=0) where in this case, $e^\nu$ is equal to $e^\beta$ and $e^\lambda$ is equal to $e^\alpha$. It can be observed on the left panel that the effect of $\kappa$ variation on $e^{\beta}$  and the difference between $e^{\beta}$ and $e^{\nu}$ appears more significant in the region closer to the center ( $r$ $\rightarrow$ 0). $e^\nu$ or  $e^\beta$ with negative  $\kappa$ value is lower than that with  positive  $\kappa$ value, while for a fixed  $\kappa$, $e^\nu$  is higher than $e^\beta$ with positive $\kappa$ and $e^\nu$  is lower than $e^\beta$ with negative $\kappa$.  However, it can be seen in the right panel that the effect of $\kappa$ variation on $e^\alpha$ appears in the region not too far from the edge. $e^\lambda$ or  $e^\alpha$ with positive  $\kappa$ value is lower than that with  negative  $\kappa$ value, while for a fixed  $\kappa$ value, the difference appears more significant in $r$  $\rightarrow$ 0 region, i.e.,  $e^\lambda$  is higher than $e^\alpha$ with positive $\kappa$ and $e^\lambda$  is lower than $e^\alpha$ with negative $\kappa$. We also need to note that in the region very close to NS radius $R$, all  $e^\alpha$ coincide with  all $e^\lambda$ while  all  $e^\beta$ coincide with  all $e^\nu$. These behaviors are due to different sign of $\kappa$ that yields small but crucial difference in region  $r$ $\rightarrow$ 0  of parameter $a$ and $b$ profiles. The plots of $a$ and $b$ profiles are shown on the left and right panels of Fig.~\ref{fig:metric_f}.

In Fig.~\ref{fig:pressure} and  Fig.~\ref{fig:eden}, the impacts of applying  the EiBI theory of gravity on the radial profile of pressure and energy density of a NS are shown. EiBI theory of gravity yields a quite different pressure profile results compared to that of GR. For a fixed value of $P_c$, different decrement of the pressure and energy as a function of radius due to $\kappa$  variation leads to different final radius value of NS. For large $P_c$, the effect of $\kappa$ variation pressure and energy profiles in general appears in almost maximum range of the radius. On the contrary, for small $P_c$, the effect of $\kappa$ variation energy profile appears more significantly only in the region near the surface of NS. Recall that for $P_c$  $\lesssim$ 50 MeV fm$^{-3}$ the hyperons do not yet appear in NS while heavier NS mass corresponds to larger  $P_c$. Therefore, it is obvious that different behavior of pressure and energy profiles for each NS mass is caused by interplay between the attractive contribution due to the presence of hyperons and additional strong repulsive contribution if $\kappa$ $\ne$ 0 depending on their operational region. 

The correlation between $\kappa$ and $M(R)/M_\odot$ with fixed $R$ can be observed from Fig.~\ref{fig:M-K} and the one between  $\kappa$ and $R$ with fixed $M(R)/M_\odot$ is from Fig.~\ref{fig:R-K}. In Fig.~\ref{fig:M-K}, we can see that in a particular value of radius, the neutron star with a larger value of $\kappa$ yields heavier mass than that with a smaller value of $\kappa$. On the other hand, it can be seen from Fig.~\ref{fig:R-K} that in a particular value of mass, a NS with larger value of $\kappa$ yields larger radius than that with smaller $\kappa$. It is interesting to see that for relatively large and fixed $R$ (in this case $R$=13 km), the correlation between $\kappa$ and $M/M_\odot$ becomes nonlinear while for relatively large and fixed $M/M_\odot$  the correlation between $\kappa$ and $R$ is also nonlinear.  As a consequence, the effect of increasing  $\kappa$ on the NS mass appears more effective in the region where NS has small $R$ and heavier mass compared to that of large $R$ and lighter mass, even though in the region of small $R$ and heavier mass, the hyperons have already softened the EOS. 

\subsection{Comparison with observations}
Here, our main focus is comparing our numerical calculations with recent observations, such as: 2 $M_\odot$ pulsar mass, NS radius constraint and  $z$ for low mass X-ray binary EXO 0748-676 NS~\cite{Cottam02} as well as the  astrophysical and cosmological constraint of  $\kappa$ \cite{Avelino12}. 

Fig.~\ref{fig:radmass} shows the mass-radius relation for NSs with hyperons for some values of $\kappa$ predicted under the EiBI theory of gravity. We could see clearly that the NS with masses around NS maximum mass and the corresponding radii predictions are very sensitive to the $\kappa$ parameter variation. While for large $R$ ($R \gtrsim$ 15 km), the  $\kappa$ parameter variation is rather marginal. Therefore, we can adjust $\kappa$ so that the NS maximum mass prediction can be larger or smaller compared to that of GR ($\kappa$=0). A larger and positive value of $\kappa$ leads to heavier maximum mass of NSs and vice versa. In our calculation, the maximum mass $M_{\rm max}$ $\ge$ $2.0~M_{\odot}$ of NS with hyperons  can be obtained if we use $\kappa$ $\gtrsim$ $4\times10^{6}~\rm{m^2}$. Note, the corresponding NS radius  of EiBI theory of gravity is also compatible with the NS radius constraint of Ref.~\cite{Steiner2010} i.e., they fit with the vertical shaded area in the case of relatively large mass region. On the other hand, from cosmological and astrophysical constraint  for a compact object that is held by gravity as reported by \cite{Avelino12}, in this case a NS with a typical radius of about $R~\sim~12~\rm{km}$ and core density larger than $\rho~\sim~56.17~\rm{MeV~fm^{-3}}$, the constraint is $\kappa~\lesssim~6\times 10^{8}~\rm{m^2}$. A tighter constraint for $\kappa$ was also reported by the author where $\kappa~\lesssim~6 \times 10^{6}~\rm{m^2}$ will yield the corresponding mass of about $M<5M_{\odot}$. Then it is clear that if we take the $2.0~M_{\odot}$ as the constraint of lower bound of $\kappa$  and combined with the upper-bound value deducted from \cite{Avelino12}, we have restricted constraint of $\kappa$ i.e., $4\times 10^{6}~\rm{m^2}~\lesssim~\kappa \lesssim~6\times 10^{6}~\rm{m^2}$.   

In Fig.~\ref{fig:redshift1} we show the effect of $\kappa$ variation on redshift $z$ as a function of NS mass. The result is also compared to the observational constraint from EXO0748-676~\cite{Cottam02}. This constraint implies that the acceptable EOS should have maximum $z$ above 0.33~\cite{Lackey2006}. It can be seen that for $\kappa$ $\gtrsim$ 0, the results are consistent with $z$ =0.35 for $M$ $\gtrsim$  1.5  $M_\odot$. These results are quite consistent with the suspected higher masses of accreting stars in X-ray binaries. 

\begin{figure}
\epsfig{figure=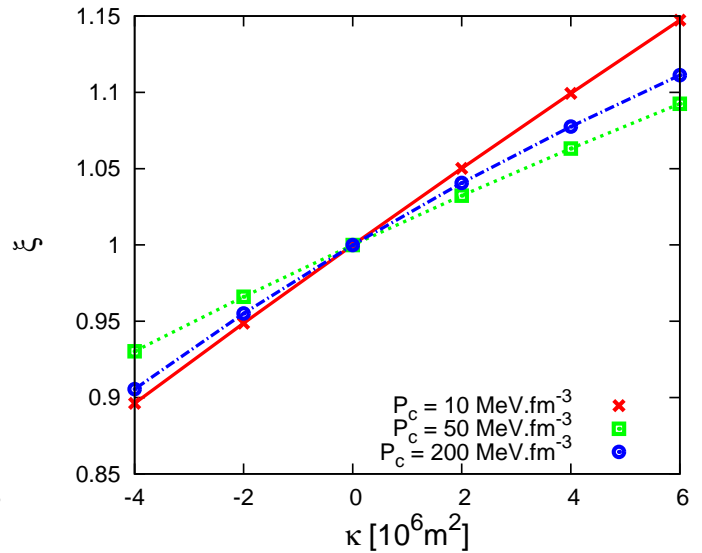, width=9.5cm}
\caption{Compactness of star  $\xi$ as a function of $\kappa$ for several fixed central pressure $P_c$.}
\label{fig:compactness}
\end{figure}

\subsection{Results on compactness}
For completeness, here we will also discuss the effect of $\kappa$ on the compactness of NS.

From Fig.~\ref{fig:compactness}, we can clearly observe that the effect of increasing $\kappa$ increases the compactness of star for a particular central pressure ($P_c$). In this work, we define the compactness of a star as $\xi =\frac{M/R}{\left({M/R}\right)_{0}}$, where $\left({M/R}\right)_{0}$ denotes the  $\left({M/R}\right)$ for $\kappa=0$ or the one that is obtained from GR. The compactness increases as $\kappa$ used in calculation increases. If we assume that the limit of causality predicted by EiBI is the same as the one of GR where the limit of causality predicted by GR~\cite{LP2007} is $R\gtrsim2.83~GM/c^2$, then we can estimate the upper limit for compactness as $\xi_{\rm{critical}}\simeq0.24~\frac{M_{\odot}}{\rm{km}}/\left({\frac{M}{R}}\right)_{0}$. So, for each $P_c$ in Fig~\ref{fig:compactness}, we can obtain $\xi_{\rm{critical}}\simeq5.56$ for $P_c$=10 MeV fm$^{-3}$,  $\xi_{\rm{critical}}\simeq2.30$ for $P_c$=50  MeV fm$^{-3}$ and $\xi_{\rm{critical}}\simeq1.66$ for  $P_c$=200  MeV fm$^{-3}$. These results are much greater than the compactness we obtain in Fig~\ref{fig:compactness} especially in the region of constraint $4\times 10^{6}~\rm{m^2}~\lesssim~\kappa \lesssim~6\times 10^{6}~\rm{m^2}$. If we compare the result in Fig.~\ref{fig:radmass}, where the M-R of the corresponding NS are below the causality limit of GR, and the result in Fig.~\ref{fig:compactness}, where the compactness increases as $\kappa$ increases, it is obvious that the  mass-radius relation of NS based on EiBI theory of gravity with $\kappa$ value in the acceptable range does not violate causality constraint. 
    
\section{CONCLUSIONS}
\label{sec_conclu}
In this work, we calculate NS mass-radius relation where hyperons are present in NS core by applying the EiBI theory of gravity. The NS core EOS with hyperons is calculated by using ERMF model where standard SU(6) prescription and hyperons potential depths  are used to determine the hyperon coupling constants while the crust EOS is taken from Ref.~\cite{MYN2013}. We have found that:
\begin{enumerate}
\item The $\kappa$ parameter of EiBI theory of gravity plays significant role in increasing or decreasing the maximum mass of NSs. This result is consistent with the one obtained by Sotani\cite{Sotani14} though the author uses different EOS to the one used in this work. If we take value for $\kappa$ around $4.0\times 10^{6}~\rm{m}^2$, the maximum mass and its corresponding radius are compatible with the constraints of Refs.\cite{Antoniadis13,Steiner2010}. Furthermore, the corresponding   $\kappa$ value is also consistent with the range predicted by  astrophysical and cosmological observations\cite{Avelino12}.
\item In NS core, the $\kappa$ parameter of EiBI theory of gravity plays a role to decrease or increase the pressure and energy density as function of radius that causes the NSs with the same central pressure larger or smaller depending on the sign of  $\kappa$ parameter.
\item Larger $\kappa$ parameter variation increases the compactness of NS and for the $\kappa$ within $4\times 10^{6}~\rm{m^2}~\lesssim~\kappa \lesssim~6\times 10^{6}~\rm{m^2}$, the corresponding mass-radius relation does not violate the causality constraint.
\end{enumerate}

Our results might indicate that NS can be an astrophysical tool to probe the deviation of gravity from Einstein's theory of General Relativity. This is a very intriguing possibility and deserves further study.\\

\section*{ACKNOWLEDGMENT}

We thank I. Prasetyo and B. A. Cahyo for useful discussions. This work has been partly supported by the Research Cluster Grant Program of the University of Indonesia, under contract No.~1709/H2.R12/HKP.05.00/2014 and No.~1862/UN.R12/HKP.05.00/2015. We acknowledge the supports provided by Universitas Indonesia.


\begin {thebibliography}{50}
\bibitem{Will2009}C. M. Will,
\Journal{Space. Sci. Rev}{148}{3}{2009}.
\bibitem{Psaltis2008}D. Psaltis,
\Journal{\LRR}{11}{9}{2008}.
\bibitem{Lattimer2012}J. M. Lattimer,
\Journal{\ARNPS}{62}{485}{2012}.
\bibitem{Chamel2013}N. Chamel, P. Haensel, J. L. Zdunik, and A. F. Fantina,
\Journal{\IJMPE}{22}{1330018}{2013}.
\bibitem{KKYT2013} B. Kiziltan, A. Kottas, M. D. Yoreo, and S. E. Thorsett,
\Journal{\AJ}{778}{66}{2013}.
\bibitem{Demorest10} P.B. Demorest, T. Pennucci, S.M. Ransom, M.S.E. Roberts , and J.W.T. Hessels, 
\Journal{Nature}{467}{1081}{2010}.
\bibitem{Antoniadis13} J. Antoniadis, {\it et al},
\Journal{Science}{340}{6131}{2013}.
\bibitem{KBK2011} M. H. van Kerkwijk, R. P. Breton, and S. R. Kulkarni,
\Journal{\AJ}{728}{95}{2011}.
\bibitem{RFSC2012}R. W. Romani,  {\it et al},
\Journal{\AJL}{760}{L36}{2012}.
\bibitem{MCM2013} M. C. Miller, arXiv:1312.0029 [astro-ph.He].
\bibitem{Bog2013}S. Bogdanov,
\Journal{\AJ}{762}{96}{2013}.
\bibitem{Gui2013}S. Guillot, M. Servillat, N. A. Webb, and R. E. Rutledge,
\Journal{\AJ}{772}{7}{2013}.
\bibitem{LS2013}J. M. Lattimer and A. W. Steiner, arXiv:1305.3242 [astro-ph.He].
\bibitem{Leahy2011}D. A. Leahy, S. M. Morsink, and Y. Chou,
\Journal{\AJ}{742}{17}{2011}.
\bibitem{Ozel2010}F.\"Ozel, G. Baym, and T. G\"uver,
\Journal{\PRD}{82}{101301}{2010}.
\bibitem{Steiner2010}A. W. Steiner, J. M. Lattimer, and E. F. Brown,
\Journal{\AJ}{722}{33}{2010}.
\bibitem{Stein2013}A. W. Steiner, J. M. Lattimer, and E. F. Brown,
\Journal{\AJL}{765}{5}{2013}.
\bibitem{Sule2011}V. Suleimanov, J. Poutanen, M. Revnivtsev, and K. Werner,
\Journal{\AJ}{742}{122}{2011}.
\bibitem{LattimerLim}J. M. Lattimer and Y. Lim,
\Journal{\AJ}{771}{51}{2013}.
\bibitem{Hyppuzz}N. K. Glendenning, S. A. Moszkowski, \Journal{\PRL}{67}{2414}{1991}; N. K. Glendenning, J. Schaffner-Bielich, \Journal{\PRL}{81}{4564}{1998}; J. Schaffner-Bielich, \Journal{\NPA}{804}{309}{2008}; I. Vidana, A. Polls, A. Ramos, L. Engvik, and M. Hjorth-Jensen, \Journal{\PRC}{62}{035801}{2000}; H.-J. Schulze and T. Rijken, \Journal{\PRC}{84}{035801}{2011}; J. L. Zdunik, and P. Haensel, \Journal{\AA}{551}{A61}{2013}.
\bibitem{Lonardoni}D. Lonardoni, A.Lovato, S. Gandolfi, and F. Pederiva,
\Journal{\PRL}{114}{092301}{2015}.
\bibitem{Yamamoto}Y. Yamamoto, T. Furumoto, N. Yasutake, and Th. A. Rijken,
\Journal{\PRC}{90}{045805}{2014}.
\bibitem{Artyom14}A. V. Astashenok, S. Capozziello, and S. D. Odintsov,
\Journal{\PRD}{89}{103509}{2014}.
\bibitem{DeDeo2003}S. DeDeo and D. Psaltis,
\Journal{\PRL}{90}{141101}{2003}.
\bibitem{Kazim2014}K. Y. Eksi, C. G\"ung\"or and M. M. T\"urko$\breve{g}$lu,
\Journal{\PRD}{89}{063003}{2014}.


\bibitem{Banados10} M. Banados and P.G. Ferreira, Phys. Rev. Lett. {\bf 105}, 011101 (2010).
\bibitem{Sham2013}Y.-H. Sham, P. T. Leung, and L.-M. Lin, Phys. Rev. D {\bf 87}, 061503 (R)(2013).
\bibitem{Kim2014}H-C. Kim, 
\Journal{\PRD}{89}{064001}{2014}.
\bibitem{Sotani14}H. Sotani, 
\Journal{\PRD}{89}{104005}{2014}.
\bibitem{Avelino12}P. P. Avelino, 
\Journal{\PRD}{85}{104053}{2012}.

\bibitem{Furnstahl96} R. J. Furnstahl, B. D. Serot, and H. B. Tang,
\Journal{\NPA}{598}{539}{1996}.
\bibitem{Furnstahl97} R. J. Furnstahl, B. D. Serot, and H. B. Tang,
\Journal{\NPA}{615}{441}{1997}.
\bibitem{SA2012} A. Sulaksono and B. K. Agrawal,
\Journal{\NPA}{895}{44}{2012}.
\bibitem{AS2014} A. Sulaksono,
\Journal{\IJMPE}{24}{1550007}{2015}.
\bibitem{MYN2013}T. Miyatsu, S. Yamamuro, and K. Nakazato,
\Journal{\AJ}{777}{4}{2013}.
\bibitem{JSB_AG} J. Schaffner-Bielich, and A. Gal,
\Journal{\PRC}{62}{034311}{2000}.


\bibitem{Glendenning} N. K. Glendenning, {\it Compact Stars: Nuclear Physics, Particle Physics, and General Relativity} (Springer-Verlag, New York, 2000), 2nd Ed.
\bibitem{Walecka} J. D. Walecka, {\it Theoretical Nuclear and Subnuclear Physics, Second Edition} (Imperial Collage Press and World Scientific Publishing, London, 2004).
\bibitem{Born:1934gh} 
  M.~Born and L.~Infeld,
  Proc.\ R.\ Soc.\ London\ {\bf A 144}, 425 (1934).
\bibitem{Harko13} T. Harko, F. S. N. Lobo, M.K. Mak, and S.V. Sushkov, 
\Journal{\PRD}{88}{044032}{2013}.
\bibitem{Delsate12} T. Delsate and J. Steinhoff, 
\Journal{\PRL}{109}{021101}{2012}.
\bibitem{Vollick04} D. N. Vollick, 
\Journal{\PRD}{69}{064030}{2004}.
\bibitem{LP2007} J. M. Lattimer and M. Prakash, 
\Journal{Phys. Rept.}{442}{109}{2007}.
\bibitem{Cottam02} J. Cottam, F. Paerels, and M. Mendez, 
\Journal{Nature}{420}{51}{2002}.
\bibitem{Lackey2006}B. D. Lackey, M. Nayyar, and B. J. Owen,
\Journal{\PRD}{73}{024021}{2006}.
\end{thebibliography}
\end{document}